\documentclass[journal]{IEEEtran}

\usepackage{setspace}
\usepackage{epsfig}  % for figures
\usepackage{graphicx}  % another package that works for figures
\usepackage{amsmath}  % for math spacing
\usepackage{amssymb}  % for math spacing
\usepackage{lscape}  % Useful for wide tables or figures.
\usepackage[justification=raggedright,font=small]{caption}	% makes captions ragged right - thanks to Bryce Lobdell
\usepackage[labelformat=simple]{subcaption}

\usepackage{color}
\usepackage{cite}
\usepackage{mathrsfs}
\usepackage{mathtools}
\usepackage{amsthm}
\usepackage{array}
\usepackage{booktabs}
\usepackage{bm}
\DeclarePairedDelimiter\abs{\lvert}{\rvert}
\newcommand{\textapprox}{\raisebox{0.5ex}{\texttildelow}}

\ifCLASSINFOpdf
\else
\fi

\begin{document}

\title{Full-Wave Methodology to Compute the Spontaneous Emission Rate of a Transmon Qubit }

\author{Thomas E. Roth~\IEEEmembership{Member,~IEEE},
        and Weng C. Chew~\IEEEmembership{Life Fellow,~IEEE}% <-this % stops a space
\thanks{This work was supported by a startup fund at Purdue University and the Distinguished Professorship Grant at Purdue University.
	
Thomas E. Roth is with the Elmore Family School of Electrical and Computer Engineering, Purdue University, West Lafayette, IN 47907 USA. Weng C. Chew is with the Department of Electrical and Computer Engineering, University of Illinois at Urbana-Champaign, Urbana, IL 61801 USA and the Elmore Family School of Electrical and Computer Engineering, Purdue University, West Lafayette, IN 47907 USA (contact e-mail: rothte@purdue.edu).

This paper is an expanded version of a conference paper presented at 2021 IEEE International Symposium on Antennas and Propagation and USNC-URSI Radio Science Meeting, Singapore.

This work has been submitted to the IEEE for possible publication. Copyright may be transferred without notice, after which this version may no longer be accessible.}% <-this % stops a space
}

%\markboth{Journal of \LaTeX\ Class Files,~Vol.~13, No.~9, September~2014}%
%{Shell \MakeLowercase{\textit{et al.}}: Bare Demo of IEEEtran.cls for Journals}

\maketitle

\begin{abstract}
The spontaneous emission rate (SER) is an important figure of merit for any quantum bit (qubit), as it can play a significant role in the control and decoherence of the qubit. As a result, accurately characterizing the SER for practical devices is an important step in the design of quantum information processing devices. Here, we specifically focus on the experimentally popular platform of a transmon qubit, which is a kind of superconducting circuit qubit. Despite the importance of understanding the SER of these qubits, it is often determined using approximate circuit models or is inferred from measurements on a fabricated device. To improve the accuracy of predictions in the design process, it is better to use full-wave numerical methods that can make a minimal number of approximations in the description of practical systems. In this work, we show how this can be done with a recently developed field-based description of transmon qubits coupled to an electromagnetic environment. We validate our model by computing the SER for devices similar to those found in the literature that have been well-characterized experimentally. We further cross-validate our results by comparing them to simplified lumped element circuit and transmission line models as appropriate.   
\end{abstract}


\IEEEpeerreviewmaketitle

\section{Introduction}
\label{sec:intro}
\IEEEPARstart{O}{f} the many hardware platforms being pursued to develop quantum information processing devices, circuit quantum electrodynamics (QED) architectures are one of the most popular due to the engineering control that is capable with these systems \cite{blais2004cavity,blais2007quantum,gu2017microwave}. Built with superconductors, these devices leverage the quantized interactions between electromagnetic fields in the microwave frequency regime and large collections of Cooper pairs (charge carriers in superconductors) on a macroscopic scale of typical circuit components. Due to the macroscopic size of these systems, the interactions with electromagnetic fields can be strongly controlled using microwave engineering principles. As a result, these systems utilize many circuit components and techniques that are very familiar to classical microwave engineers; including coplanar waveguide transmission lines and resonators, circulators, wire bonds or airbridges, and distributed inductors and capacitors, to name a few \cite{gu2017microwave,krantz2019quantum,ranzani2019circulators}. Using these tools, circuit QED designs for a wide range of quantum technologies have been demonstrated, including analog quantum computers \cite{ma2019dissipatively}, digital or gate-based quantum computers \cite{arute2019quantum,wu2021strong,kandala2017hardware,barends2014superconducting}, single photon sources \cite{houck2007generating,zhou2020tunable,lang2013correlations}, quantum memories \cite{reagor2016quantum,sardashti2020voltage}, and components of quantum communication systems \cite{leung2019deterministic}.

Due to the wide variety of circuit QED  technologies being developed, the figures of merit that need to be characterized in the design process are similarly broad. However, one characteristic that is important for all of these quantum technologies is the spontaneous emission rate (SER) of the artificial atom that serves as the quantum bit (qubit) in these devices \cite{houck2008controlling}. Depending on other device design and fabrication factors, the SER of the qubit can play a significant role in the overall relaxation time (denoted as $T_1$) of the qubit. 

It is well known that the SER of a qubit depends not only on the internal properties of the qubit, but also on the structure of the electromagnetic environment the qubit is coupled to (this is known as the Purcell effect) \cite{purcell1946resonance}. Hence, engineering the electromagnetic environment around a qubit can provide a vital tool for controlling the relaxation time of a qubit. This is imperative for many quantum technologies; e.g., in single photon sources where it is desirable to enhance the SER to make the source ``on-demand'' \cite{houck2007generating,zhou2020tunable,lang2013correlations} or in quantum computers where the SER must be suppressed to not limit qubit coherence \cite{houck2008controlling}. As a result, there is a need for accurate numerical predictions of the SER of a qubit over many orders of magnitude and possible operating frequencies (especially in the case of frequency-tunable qubits).

Traditionally, the SER of a superconducting qubit has been considered from a lumped element circuit theory perspective \cite{esteve1986effect,neeley2008transformed}. Here, it was found that when spontaneous emission dominated the qubit lifetime, $T_1$ was approximately equal to the classical decay time of the circuit. This result gave $T_1 \approx C_q/\mathrm{Re}\{Y_\mathrm{eq}(\omega)\}$, with $C_q$ being the qubit capacitance and $Y_\mathrm{eq}(\omega)$ the equivalent admittance of the entire circuit seen by the qubit \cite{neeley2008transformed}. Hence, many tools from \textit{classical} circuit and microwave engineering disciplines can be used to control this important \textit{quantum} property of a superconducting qubit. Although convenient in its simplicity, this result was found to lead to significant inaccuracy (multiple orders of magnitude) if $Y_\mathrm{eq}(\omega)$ was not computed accurately enough by including a sufficient number of modes of the electromagnetic system (e.g., a transmission line resonator) in the model  \cite{houck2008controlling}. This issue was overcome in \cite{houck2008controlling} by extending the circuit theory result to consider a more complete transmission line model for the calculation of $Y_\mathrm{eq}(\omega)$.

Obviously, any circuit or transmission line theory approach necessarily incorporates multiple approximations of varying levels of accuracy to the complete electromagnetic physics involved depending on the system being analyzed. As a result, it is desirable to have a general full-wave theory to characterize the SER of superconducting qubits. This provides a method for attaining the utmost modeling accuracy, and can also guide the development of simpler circuit and transmission line models for complex systems, as appropriate. 

In this work, we develop such a full-wave theory for characterizing the SER of one of the most popular superconducting qubits; namely, the transmon qubit \cite{koch2007charge,roth2021introduction}. This qubit is popular due to its resilience to certain common sources of decoherence and the maturity of qubit control and readout techniques; leading to it being a key component in recent demonstrations of a computational quantum advantage \cite{arute2019quantum,wu2021strong}. To facilitate the development of a full-wave theory for the SER of a transmon qubit, we use our recently developed field-based description of circuit QED systems involving transmon qubits \cite{roth2021macroscopic} and other standard results from QED theory \cite{novotny2012principles}.

Preliminary results on the development of a full-wave theory for computing the SER of transmon qubits were reported in \cite{roth2021full}. This work expands on \cite{roth2021full} by providing more details on the theoretical derivation and by further extending the derivation to highlight how standard computational electromagnetics methods can be used to easily perform the needed analysis. Further, we present many new numerical examples that are designed to closely replicate devices characterized experimentally in the literature to provide qualitative validation of our approach. 

The remainder of this work is organized in the following way. In Section \ref{sec:background}, we review the necessary background on the field-based description of circuit QED architectures with transmon qubits developed in \cite{roth2021macroscopic}. Following this, we present in Section \ref{sec:formulation} the formulation of how this field-based description can be used to develop a full-wave computation process for the SER of transmon qubits. We then validate our formulation by comparing numerical results to various experimental and theoretical results in Section \ref{sec:results}. Finally, we present conclusions on this work in Section \ref{sec:conclusion}.

\section{Field-Based Circuit QED Background}
\label{sec:background}
Before presenting the full-wave approach to computing the SER of transmon qubits, it is necessary to review certain details of the field-based description of circuit QED systems involving transmon qubits developed in \cite{roth2021macroscopic}. There, it was shown that the Hamiltonian describing the coupled field-transmon system could be given as
\begin{align}
	\hat{H} = \hat{H}_T + \hat{H}_F + \hat{H}_I,
\end{align}
where $\hat{H}_{T}$ is the free transmon Hamiltonian, $\hat{H}_F$ is the free field Hamiltonian, and $\hat{H}_I$ is the interaction Hamiltonian describing the coupling between the two systems. 

More specifically, we have for the free transmon Hamiltonian that 
\begin{align}
	\hat{H}_T = 4 E_C \hat{n}^2 - E_J \cos\hat{\varphi},
	\label{eq:free-transmon}
\end{align}
where $\hat{n}$ is the charge operator of the transmon qubit that counts the number of Cooper pairs that have tunneled through the Josephson junction of the qubit compared to some equilibrium reference point and $\hat{\varphi}$ is the phase operator that characterizes the phase difference across the Josephson junction of the transmon qubit \cite{koch2007charge,roth2021introduction}. We further have that $E_C$ is the single electron charging energy of the total capacitance of the qubit system, and that $E_J$ is the Josephson energy that measures the energy associated with a Cooper pair tunneling through the junction. Transmon qubits are often characterized as an optimized form of charge qubit with $E_J/E_C \gg 1$ \cite{koch2007charge,roth2021introduction}.

Next, we have for the free field Hamiltonian that 
\begin{align}
	\hat{H}_F = \frac{1}{2}  \iiint \big( \epsilon \hat{\mathbf{E}}^2 + \mu \hat{\mathbf{H}}^2 \big) d\mathbf{r},
	\label{eq:free-field}
\end{align}
where $\hat{\mathbf{E}}$ and $\hat{\mathbf{H}}$ are the electric and magnetic field operators, respectively. This free field Hamiltonian can be easily recognized as the total electromagnetic energy within a volume by comparing it to the well-known Poynting's theorem. The particular mathematical expressions of $\hat{\mathbf{E}}$ and $\hat{\mathbf{H}}$ depend on a number of choices in the quantization approach used, with a few relevant options discussed in detail in \cite{roth2021macroscopic}. We will discuss a simple form for these operators relevant to this work shortly.

Finally, we have for the interaction Hamiltonian that
\begin{align}
	\hat{H}_I = - \iiint  \hat{\mathbf{E}} \cdot \partial_t^{-1} \hat{\mathbf{J}}_T  d\mathbf{r},
	\label{eq:int-H}
\end{align} 
which describes the coupling between the two systems in terms of $\hat{\mathbf{E} }$ and a transmon current density operator $\hat{\mathbf{J}}_T$. The transmon current density operator is
\begin{align}
	\hat{\mathbf{J}}_T = -2e  \mathbf{d}(\mathbf{r}) \partial_t\hat{n},
	\label{eq:transmon-current-operator1}
\end{align}
where $e$ is the electron charge. We use the somewhat awkward notation of $\partial_t^{-1}\hat{\mathbf{J}}_T$ in (\ref{eq:int-H}) for consistency with \cite{roth2021macroscopic}, where it was more convenient to use this form in deriving equations of motion. We will simplify the expressions later to keep the notation more compact in this work. Beyond this detail, in (\ref{eq:transmon-current-operator1}), $\mathbf{d}$ is a vector characterizing the line integration path that would define the voltage seen by the Josephson junction in the transmon qubit. Physically, we can see that this operator considers the changes in the number of Cooper pairs that have tunneled through the Josephson junction. This naturally produces a current, making the designation of this operator as a current density physically intuitive.

The purpose of the definitions in (\ref{eq:transmon-current-operator1}) is to have the volume integration in (\ref{eq:int-H}) reduce to the evaluation of the voltage due to $\hat{\mathbf{E}}$ at the location of the transmon. The exact form of $\mathbf{d}$ will depend on the particular transmon geometry considered and how it is modeled. By selecting our definitions in this way, it can be shown that after adopting standard transmission line and lumped element circuit approximations this field-based Hamiltonian reduces to the typical Hamiltonian used to study transmon systems in the literature \cite{roth2021macroscopic}. 
\section{Formulation}
\label{sec:formulation}
With the necessary background in place, we may now formulate how to compute the SER of a transmon qubit using full-wave methods. We begin in Section \ref{subsec:spont-emission-QED} by showing how the results from the field-based description of the system can be used to express the SER of the transmon qubit in terms of the dyadic Green's function of the electromagnetic system. Following this, we discuss in Section \ref{subsec:full-wave-comp} how standard full-wave methods can be used to efficiently compute the effect of the dyadic Green's function needed in the evaluation of the SER of the transmon.

\subsection{QED of Spontaneous Emission for the Transmon Qubit}
\label{subsec:spont-emission-QED}
In the weak-coupling regime of cavity QED, the transition rate between two states of a quantum system can be analyzed using a relatively simple time-dependent perturbation theory analysis \cite{miller2008quantum}. The result, often referred to as Fermi's golden rule, can be used to compute the SER of a qubit by selecting appropriate initial and final states of the system to compute the transition rate between. Although this result is only applicable to the ``weak-coupling'' regime of cavity QED, this is the relevant experimental regime for many current quantum technologies.

Considering this, Fermi's golden rule gives the SER at frequency $\omega_0$ between an initial state $|i\rangle$ and a final state $|f\rangle$ as
\begin{align}
	\gamma_{(f,i)}(\omega_0) =  \frac{2\pi}{\hbar^2} \abs{ \langle f | \hat{H}_I | i \rangle   }^2 \, \delta(\omega_{fi}-\omega_0),
	\label{eq:fermi-golden}
\end{align}
where $\hat{H}_I$ is the interaction Hamiltonian between the qubit and the electromagnetic field and $\omega_{fi}$ is the frequency associated with the energy difference between states $|i\rangle$ and $|f\rangle$ \cite{novotny2012principles}. To evaluate $\abs{ \langle f | \hat{H}_I | i \rangle   }^2$ in a computationally convenient manner, it is necessary to provide a more explicit expression for $\hat{\mathbf{E}}$ in the interaction Hamiltonian given in (\ref{eq:int-H}). As we will see shortly, an efficient computational process can be developed by first considering a mode decomposition representation of $\hat{\mathbf{E}}$. This then allows for the evaluation of $\abs{ \langle f | \hat{H}_I | i \rangle   }^2$ to be related to the dyadic Green's function of the electromagnetic system the qubit is embedded in \cite{novotny2012principles}, which can be computed efficiently using computational electromagnetics methods.

The particular mathematical description of $\hat{\mathbf{E}}$ depends on the quantization process used. Here, we follow a simple mode decomposition approach discussed in detail in \cite{roth2021macroscopic}. We consider our electromagnetic system to contain inhomogeneous, lossless, and non-dispersive dielectric and perfectly conducting regions only. We further consider the quantization of the electromagnetic field within a macroscopic QED framework \cite{scheel2008macroscopic}. The key aspect of this is that a microscopic description of a lossless, non-dispersive dielectric medium is not needed. Instead, macroscopic permittivities and permeabilities may be used directly in the quantum description of the electromagnetic fields in the same way they are used for a classical description.

To keep the notation simple, we will only consider quantizing the electromagnetic field for a discrete spectrum of modes. This implies that our analysis is to be considered in a closed region, such as within a perfectly conducting cavity \cite{roth2021macroscopic}. Although this is not truly the case for practical devices, the final result of this derivation in terms of the dyadic Green's function is still applicable to open region problems. We will discuss this point in more detail at the end of this section.

Now, within a closed system, we can use a separation of variables argument to write the classical electric field as
\begin{align}
	\mathbf{E}(\mathbf{r},t) = \sum_k \sqrt{\frac{\omega_k}{2\epsilon_0}} \big( q_k(t) + q_k^*(t)  \big) \mathbf{E}_k(\mathbf{r}) ,
\end{align}
where $\omega_k$ is the eigenvalue of the $k$th mode. We can insert this representation into the wave equation,
\begin{align}
	\nabla\times\nabla\times\mathbf{E}(\mathbf{r},t) + \mu \epsilon(\mathbf{r}) \partial_t^2 \mathbf{E}(\mathbf{r},t) = 0,
\end{align}
to find two separated equations for each mode, given by
\begin{align}
	\partial_t^2 q_k(t) = -\omega_k^2 q_k(t),
	\label{eq:sep-q}
\end{align}
\begin{align}
	\nabla\times\nabla\times\mathbf{E}_k(\mathbf{r}) - \mu\epsilon(\mathbf{r}) \omega_k^2 \mathbf{E}_k(\mathbf{r}) = 0.
	\label{eq:sep-E}
\end{align}
We further require that for the eigenvalue problem given by (\ref{eq:sep-E}) the modes be orthonormal such that
\begin{align}
	\iiint \epsilon_r(\mathbf{r}) \mathbf{E}^*_{k_1}(\mathbf{r}) \cdot \mathbf{E}_{k_2}(\mathbf{r}) d\mathbf{r} = \delta_{k_1,k_2},
\end{align}
where $\delta_{k_1,k_2}$ is the Kronecker delta function. A similar mode expansion also holds for the magnetic field \cite{roth2021macroscopic}.

These modal expansions can be substituted into the free field Hamiltonian given in (\ref{eq:free-field}) to find that each mode behaves like an uncoupled harmonic oscillator, or equivalently like uncoupled LC resonant circuits. As a result, a canonical quantization process can be performed and bosonic annihilation and creation operators can be introduced for each mode, denoted as $\hat{a}_k$ and $\hat{a}^\dagger_k$, respectively \cite{roth2021macroscopic,chew2016quantum2}. The final result is that we can write the electromagnetic field operator $\hat{\mathbf{E}}$ as
\begin{align}
	\hat{\mathbf{E}}(\mathbf{r},t) = \sum_k \sqrt{\frac{\hbar \omega_k}{2\epsilon_0}} \big(  \hat{a}_k(t) + \hat{a}^\dagger_k(t) \big) \mathbf{E}_k(\mathbf{r}).
	\label{eq:quant-mode-exp}
\end{align}
As mentioned previously, this expression is valid for closed regions with ``standing wave'' modes. Expressions for more general situations can be found in \cite{roth2021macroscopic}.

We may now substitute (\ref{eq:quant-mode-exp}) into the interaction Hamiltonian given in (\ref{eq:int-H}) to get
\begin{align}
	\hat{H}_I =  \sum_k 2e\sqrt{\frac{\hbar \omega_k}{2\epsilon_0}} \big( \hat{a}_k + \hat{a}^\dagger_k \big) \hat{n} \iiint \mathbf{E}_k(\mathbf{r}) \cdot \mathbf{d}(\mathbf{r}) \, d\mathbf{r},
	\label{eq:int-H2}
\end{align}
after expanding all definitions for the various operators out explicitly. Next, we can express $\hat{n}$ in terms of eigenstates $|j\rangle$ of the free transmon Hamiltonian (\ref{eq:free-transmon}) using the resolution of the identity operator \cite{ryu2021fourier}. This gives
\begin{align}
	\hat{n} = \sum_j \langle j | \hat{n} | j+1 \rangle \big[ |j\rangle \langle j+1 | + \mathrm{H.c.}   \big] ,
\end{align}
where we have also used the result that $\hat{n}$ only couples nearest-neighbor eigenstates in the transmon operating regime and that $ \langle j | \hat{n} | j+1 \rangle =  \langle j+1 | \hat{n} | j \rangle$ \cite{koch2007charge}. Substituting this into (\ref{eq:int-H2}) and applying the rotating wave approximation (which is valid in the operating regimes where Fermi's golden rule is applicable), yields
\begin{multline}
	\hat{H}_I = \sum_{k,j}2e \langle j | \hat{n} | j+1 \rangle \sqrt{\frac{\hbar\omega_k}{2\epsilon_0}} \iiint \mathbf{E}_k(\mathbf{r}) \cdot \mathbf{d}(\mathbf{r}) \, d\mathbf{r} \\ \times \bigg[ \hat{a}_k^\dagger |j\rangle\langle j+1| + \mathrm{H.c.}  \bigg].
	\label{eq:int-H3}
\end{multline} 

To proceed, we now need to specify the states which we want to compute the SER between. As an initial state, we consider the transmon to be in some excited state and the electromagnetic system to be in its vacuum state. This joint state is denoted as $|j+1,\{0\}\rangle$. The final states that we compute the transition rate to then have the transmon in its next lowest state and all single photon states that are resonant with the energy difference between $|j\rangle$ and $|j+1\rangle$ (i.e., the photon states have frequency equal to $\omega_0 = (E_{j+1}-E_{j})/\hbar$, where $E_j$ is the energy associated with state $|j\rangle$). These joint states are denoted as $|j,\{1_k\}\rangle$. 

Focusing on these states, we can substitute (\ref{eq:int-H3}) into Fermi's golden rule given in (\ref{eq:fermi-golden}) to get
\begin{multline}
	\gamma_{(f,i)}(\omega_0) = \frac{\pi \omega_0}{\hbar \epsilon_0} \big( 2e \abs{ \langle j | \hat{n} | j+1\rangle }  \big)^2 \delta(\omega_k-\omega_0) \\ \times \sum_k \iint \mathbf{d}(\mathbf{r}) \cdot \mathbf{E}_k(\mathbf{r}) \mathbf{E}^*_k(\mathbf{r}') \cdot \mathbf{d}(\mathbf{r}') \, d\mathbf{r}' d\mathbf{r},
	\label{eq:ser2}
\end{multline}
where the delta function has been rewritten to enforce that the field modes are resonant with the frequency the SER is being evaluated at. The form of the eigenmode expansion in (\ref{eq:ser2}) is similar to that seen in the eigenmode expansions of a dyadic Green's function \cite{novotny2012principles}. In fact, using standard mathematical identities \cite{novotny2012principles}, it can be shown that
\begin{align}
	\mathrm{Im}\big\{ \overline{\mathbf{G}}(\mathbf{r},\mathbf{r}',\omega) \big\} = \frac{\pi c^2}{2\omega} \sum_k \mathbf{E}_k(\mathbf{r}) \mathbf{E}^*_k(\mathbf{r}') \delta(\omega-\omega_k),
\end{align}
where $\overline{\mathbf{G}}(\mathbf{r},\mathbf{r}',\omega)$ is the dyadic Green's function for the electromagnetic system. Using this in (\ref{eq:ser2}), we arrive at the desired result for the SER as
\begin{multline}
	\gamma_{(f,i)}(\omega_0) = \frac{2 \omega_0^2}{\hbar \epsilon_0 c^2} \big( 2e \abs{ \langle j | \hat{n} | j+1\rangle }  \big)^2 \\
	\times \iint \mathbf{d}(\mathbf{r})\cdot \mathrm{Im}\big\{ \overline{\mathbf{G}}(\mathbf{r},\mathbf{r}',\omega_0) \big\} \cdot \mathbf{d}(\mathbf{r}') \, d\mathbf{r}' d\mathbf{r}.
	\label{eq:ser3}
\end{multline}
Typically, the SER of a quantum emitter is expressed in terms of the local density of states that is proportional to $\mathrm{Im}\big\{ \overline{\mathbf{G}}(\mathbf{r}_0,\mathbf{r}_0,\omega_0) \big\}$, where $\mathbf{r}_0$ is the location of the emitter \cite{novotny2012principles,qiao2011systematic,chen2012study}. Our result in (\ref{eq:ser3}) that involves spatial integrals over the dyadic Green's function represents a generalization to this result, which can be seen to be similar to other theories for the SER of quantum emitters beyond the dipole approximation \cite{stobbe2012spontaneous}. We will discuss a simple way to compute the results of the integrals in (\ref{eq:ser3}) in Section \ref{subsec:full-wave-comp}.

Before moving on, it is necessary to comment on the assumption that the analysis is performed in a closed region. For most cases, the transmon qubits are closely coupled to nearby transmission line geometries so that it is not difficult to envision enclosing the entire device in a large box with perfectly conducting or periodic boundary conditions without significantly impacting the results. This leads to a discrete mode spectrum for the electromagnetic fields. However, modeling this large box in a full-wave tool is inconvenient and can lead to spurious results if the transmon couples to some of the cavity modes. To avoid these numerical artifacts, it is preferred to use a simple radiation boundary condition placed an appropriate distance away from the device in a full-wave model. Although the derivation of (\ref{eq:ser3}) in this section is not directly applicable to this case, (\ref{eq:ser3}) is still valid in this scenario. We have verified this through various numerical experiments, but do not show this for brevity.

A similar issue also arises for the use of microwave network ports to terminate transmission line structures in a full-wave model. In principle, this leads to a continuous mode spectrum that is more difficult to work with from a theoretical perspective (see, e.g., \cite{roth2021macroscopic}). However, so long as the dyadic Green's function used in (\ref{eq:ser3}) accounts for the presence of the ports, (\ref{eq:ser3}) is still valid for this case. In this work, we use lumped ports within a finite element method solver to act as resistive terminations to the various transmission line structures in our devices. As will be discussed in Section \ref{sec:results}, this leads to good qualitative agreement with experimental results and other approximate theoretical models.

\subsection{Full-Wave Computation}
\label{subsec:full-wave-comp}
We now turn our attention to efficiently computing the spatial integrals of the imaginary part of the dyadic Green's function needed in evaluating (\ref{eq:ser3}). It is of course possible to use computational electromagnetics methods to compute components of the dyadic Green's function through the solution of near-field scattering problems \cite{qiao2011systematic,chen2012study,rodriguez2007virtual,xiong2010efficient}. However, doing this over a sequence of points defined by the integration path $\mathbf{d}$ to numerically evaluate the spatial integrals in (\ref{eq:ser3}) is inconvenient. Instead, we can use standard field and transmission line theory results to find an alternative approach that is much simpler to implement numerically.  

To begin, we note that in the frequency domain the inhomogeneous wave equation for the classical electric field is 
\begin{align}
	\nabla\times\nabla\times\mathbf{E}(\mathbf{r},\omega) - \mu_0 \epsilon(\mathbf{r}) \omega^2 \mathbf{E}(\mathbf{r},\omega) = i\omega\mu_0 \mathbf{J}(\mathbf{r},\omega),
\end{align} 
where $\mathbf{J}$ is an impressed electric current source. We also have that the Green's function satisfies 
\begin{align}
	\nabla\times\nabla\times\overline{\mathbf{G}}(\mathbf{r},\mathbf{r}',\omega) - \mu_0 \epsilon(\mathbf{r}) \overline{\mathbf{G}}(\mathbf{r},\mathbf{r}',\omega) = \overline{\mathbf{I}} \delta(\mathbf{r}-\mathbf{r}'),
\end{align}
where $\overline{\mathbf{I}}$ is the identity dyad. Using these two results, it is easy to show that the Green's function can be used to establish a field-source relation as
\begin{align}
	\mathbf{E}(\mathbf{r},\omega) = i\omega \mu_0 \int \overline{\mathbf{G}}(\mathbf{r},\mathbf{r}',\omega) \cdot \mathbf{J}(\mathbf{r}',\omega) d\mathbf{r}'.
	\label{eq:greens1}
\end{align}

To proceed, we assume that the impressed current density can be defined so that it follows the needed line integration path in (\ref{eq:ser3}). That is, $\mathbf{J}(\mathbf{r}') = -\mathbf{d}(\mathbf{r}') I_t$, where $I_t$ is amplitude of the impressed current and the negative sign is to simplify the definitions of impedances later. Substituting this into (\ref{eq:greens1}) and taking the line integral with respect to path $\mathbf{d}(\mathbf{r})$, we get that
\begin{multline}
	 \iint \mathbf{d}(\mathbf{r})\cdot \overline{\mathbf{G}}(\mathbf{r},\mathbf{r}',\omega)  \cdot \mathbf{d}(\mathbf{r}') \, d\mathbf{r}' d\mathbf{r} \\ = \frac{i}{\omega \mu_0 I_t} \int \mathbf{d}(\mathbf{r})\cdot \mathbf{E}(\mathbf{r}) \, d\mathbf{r}.
	 \label{eq:greens2}
\end{multline}
Recalling the definition of $\mathbf{d}$ given in Section \ref{sec:background}, we can see that the integral on the right-hand side of (\ref{eq:greens2}) defines the voltage seen by the Josephson junction within the transmon qubit. Denoting this voltage as $V_t(\omega)$, we see that the right-hand side of (\ref{eq:greens2}) is proportional to an impedance $Z_t(\omega) = V_t(\omega)/I_t(\omega)$. Considering this system as a one-port network, we see that $I_t$ is the current driven through the port and $V_t$ is the corresponding port voltage. Hence, we can recognize $Z_t$ as the input impedance of this one-port network, denoted as $Z_\mathrm{in}$ \cite{pozar2009microwave}. 

Using this result in (\ref{eq:greens2}) and taking the imaginary part of both sides, we find that
\begin{align}
	\iint \mathbf{d}(\mathbf{r})\cdot \mathrm{Im}\big\{ \overline{\mathbf{G}}(\mathbf{r},\mathbf{r}',\omega) \big\} \cdot \mathbf{d}(\mathbf{r}') \, d\mathbf{r}' d\mathbf{r} = \frac{\mathrm{Re}\big\{ Z_\mathrm{in}(\omega) \}}{\omega\mu_0},
\end{align}
where we have noted that $\mathrm{Im}\{ iZ_\mathrm{in} \} = \mathrm{Re}\{ Z_\mathrm{in} \}$ to simplify the notation. The input impedance seen by the Josephson junction in the transmon can be computed easily using many different computational electromagnetics methods, and hence, leads to a much simpler procedure than directly evaluating the dyadic Green's function. With this, our final result for the SER of a transmon qubit that can be easily computed using full-wave tools is
\begin{align}
	\gamma_{(f,i)}(\omega_0) = \frac{2 \omega_0}{\hbar \mu_0 \epsilon_0 c^2} \big( 2e \abs{ \langle j | \hat{n} | j+1\rangle }  \big)^2  \, \mathrm{Re}\big\{ Z_\mathrm{in}(\omega_0) \}.
	\label{eq:ser4}
\end{align}
This result highlights that the SER is heavily influenced by the effective ``loss'' of the electromagnetic system from the perspective of the qubit, which is measured by $\mathrm{Re}\big\{ Z_\mathrm{in} \} $. The purely \textit{classical} quantity $Z_\mathrm{in}$ is very familiar to microwave engineers, who have developed sophisticated techniques that can be used to optimize it for the purposes of controlling the SER of transmon qubits. We will consider a few simple examples of ways that $\mathrm{Re}\big\{ Z_\mathrm{in} \} $ has been controlled in experiments using classical microwave design techniques in Section \ref{sec:results}.
\section{Numerical Results}
\label{sec:results}
In this section, we present the results from a number of numerical examples to test the validity of the full-wave formulation developed in this work. Unfortunately, the lack of exact analytical solutions for realistic devices and the incomplete design information available for experimentally realized devices precludes performing quantitative validation of our approach at this time. However, we can provide useful qualitative validation by designing multiple devices similar to those in the literature that exhibit significantly different SER characteristics. In all cases, we find that our numerical models provide correct trends and only differ from experimental results by a factor of $O(1)$ despite the numerous differences in our model parameters compared to fabricated devices. 

For ease of comparing our results to those available in the literature, we plot the relaxation rate of the transmon qubit rather than the SER. The correspondence is simply that the relaxation rate is given by $T_1 = 1/\gamma_{(f,i)}$. This simple equality for the total $T_1$ time only holds when no other relaxation mechanisms are considered, as is done in this work. As further cross-validation, we also compare our full-wave results to simpler lumped element circuit or transmission line modeling results where appropriate. In these cases, we compute the $T_1$ time using $T_1 = C_q/\mathrm{Re}\{Y_\mathrm{eq}\}$, where $C_q$ is the qubit capacitance and $Y_\mathrm{eq}$ is the equivalent admittance of the entire circuit as seen by the qubit \cite{houck2008controlling}. We hand tune a small number of parameters in these simpler models to achieve a good fit with our full-wave results.

Before discussing the different models studied in this work, we first comment on the evaluation of the matrix elements of the charge operator $\abs{\langle j| \hat{n} | j+1 \rangle}$ that is needed in (\ref{eq:ser4}). This may be evaluated exactly using the methods discussed in \cite{koch2007charge}. Here, we opt to use the simpler asymptotic result for the transmon that
\begin{align}
	\abs{\langle j| \hat{n} | j+1 \rangle} \approx \sqrt{\frac{j+1}{2}}\bigg(\frac{E_J}{8E_C}\bigg)^{1/4},
\end{align}
where $E_J$ is the Josephson energy, $E_C = e^2/2C_\Sigma$, and $C_\Sigma$ is the total capacitance to ground seen by the Josephson junction \cite{koch2007charge}. 

For all results presented in this work, we estimate $C_\Sigma$ from the imaginary part of the input impedance at the location of the Josephson junction. We then use the approximate asymptotic result that the operating frequency of the transmon is given by $\sqrt{8E_J E_C}/h$ to compute $E_J$ \cite{koch2007charge}. This approximately accounts for the variation in the matrix element of the charge operator as a function of frequency in the computation of the SER. However, this does at times require varying $E_J$ over unrealistically large ranges compared to what is possible physically to compare to results available in the literature. Since $E_J$ only enters the final computation with a square root dependence, this has a relatively weak effect on the overall results, and so we do not comment on this further.

\subsection{Waveguide QED Single Photon Source}
\label{subsec:waveguide-qed-sps}
\begin{figure}[t!]
	\centering
	\includegraphics[width=\linewidth]{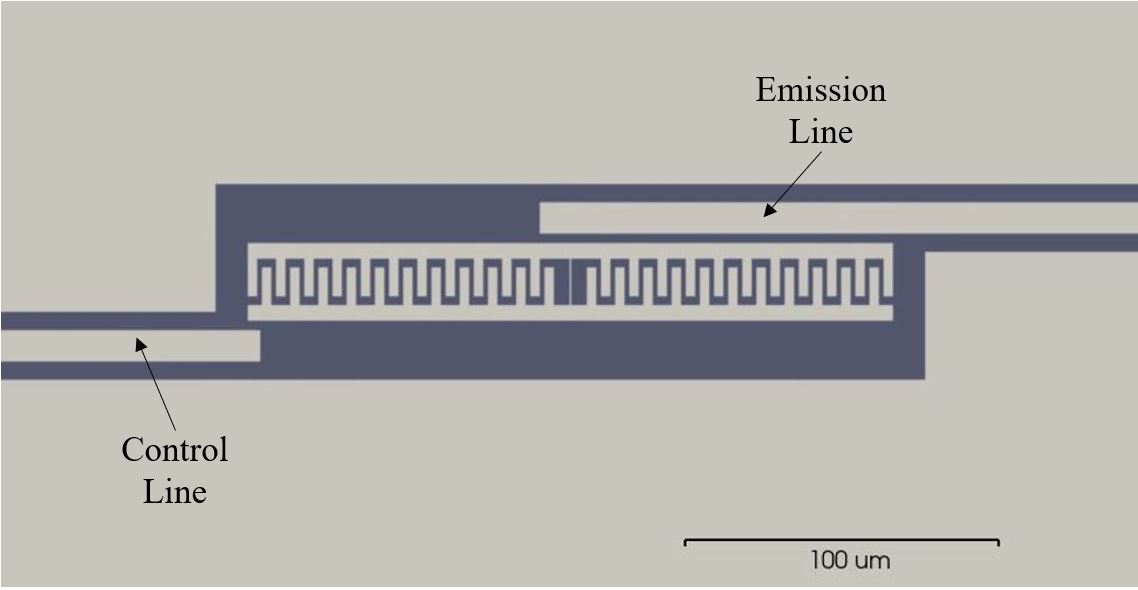}
	\caption{Waveguide QED style single photon source geometry.}
	\label{fig:zhou_sps}
\end{figure}

The first device that we will consider is a waveguide QED style single photon source that was originally presented in \cite{zhou2020tunable}. This device consists of a transmon that is capacitively coupled to two CPW lines as shown in Fig. \ref{fig:zhou_sps}. The weakly coupled line is the control line, from which classical microwave drive pulses can be applied to the transmon to control its state. This microwave drive is designed to raise the transmon with high probability into its first excited state. The transmon is then allowed to freely evolve in time until it spontaneously emits the excitation. This excitation is preferentially emitted into the second line that the transmon is more strongly coupled to, which is denoted as the emission line in Fig. \ref{fig:zhou_sps}. 

For this device, the CPW signal trace width is 10 $\mu$m and the gap width between the signal trace and ground plane is 5.8 $\mu$m. This leads to a characteristic impedance of $50 \, \Omega$ and an $\epsilon_\mathrm{eff} = 6.34$ for a thick silicon substrate. To simplify the full-wave model, we approximate the conductors as being infinitely thin and replace the typically thick substrate (usually \textapprox0.5 mm \cite{goppl2008coplanar}) with a homogeneous background relative permittivity equal to $\epsilon_\mathrm{eff}$.

Since there are no resonators involved in this device, the frequency dependence of the SER is expected to follow the simple characteristics of radiation into a continuum. From a lumped element circuit modeling perspective, this would lead to a SER of 
\begin{align}
	\gamma \approx \omega^2 Z_0 \frac{C_g^2}{C_q},
	\label{eq:approx-ser}
\end{align}
where $C_g$ is the capacitance between the transmon to the emission line and $C_q$ is the qubit capacitance \cite{houck2008controlling}. To arrive at this simplified result, the presence of the control line has been ignored in the circuit model and we have used that $(\omega C_g Z_0)^2 \ll 1$ to simplify the equivalent admittance of the circuit from the perspective of the qubit. Due to the simplicity of this device, this approximate result is expected to characterize the performance well.

\begin{figure}[t!]
	\centering
	\includegraphics[width=\linewidth]{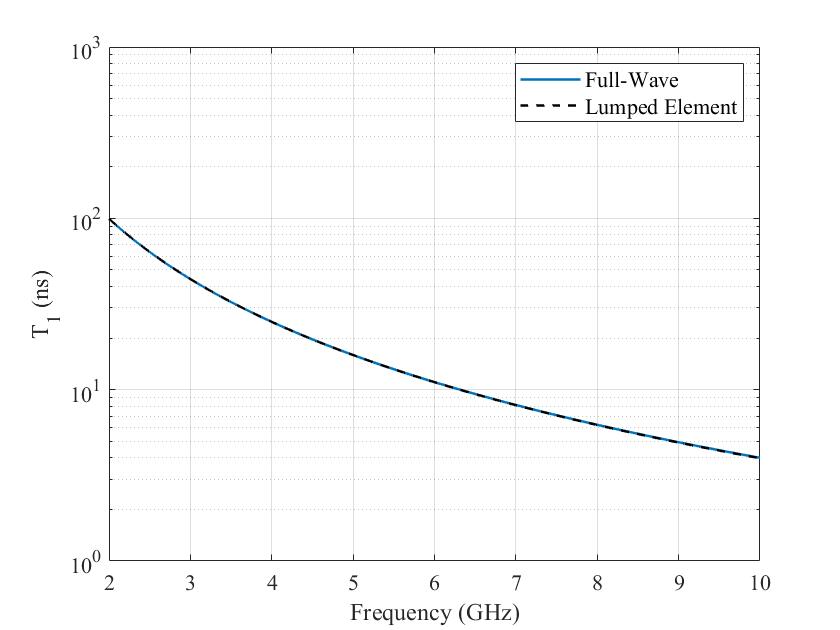}
	\caption{Predictions of the waveguide QED style single photon source $T_1$ time. Full-wave model results are shown with solid lines and lumped element model results are shown with dashed lines.}
	\label{fig:zhou_T1}
\end{figure}

We perform the comparison between the results of (\ref{eq:approx-ser}) and the full-wave analysis of the device in Fig. \ref{fig:zhou_T1}, where it is clearly seen that excellent agreement is achieved. Note that because our full-wave analysis only gives easy access to the total capacitance to ground ($C_\Sigma$), we numerically tune the ratio between $C_g$ and $C_q$ in (\ref{eq:approx-ser}) with the constraint that $C_\Sigma = C_g + C_q$ to maximize the agreement with the full-wave results. This allows us to verify the agreement in trends between the full-wave and approximate results as a function of frequency, but is obviously a limitation to the predictive power of (\ref{eq:approx-ser}) without being augmented by other numerical analysis to compute $C_g$ and $C_q$ directly.

\subsection{Cavity QED Single Photon Source}
\label{subsec:cavity-qed-sps}
\begin{figure}[t!]
	\centering
	\includegraphics[width=\linewidth]{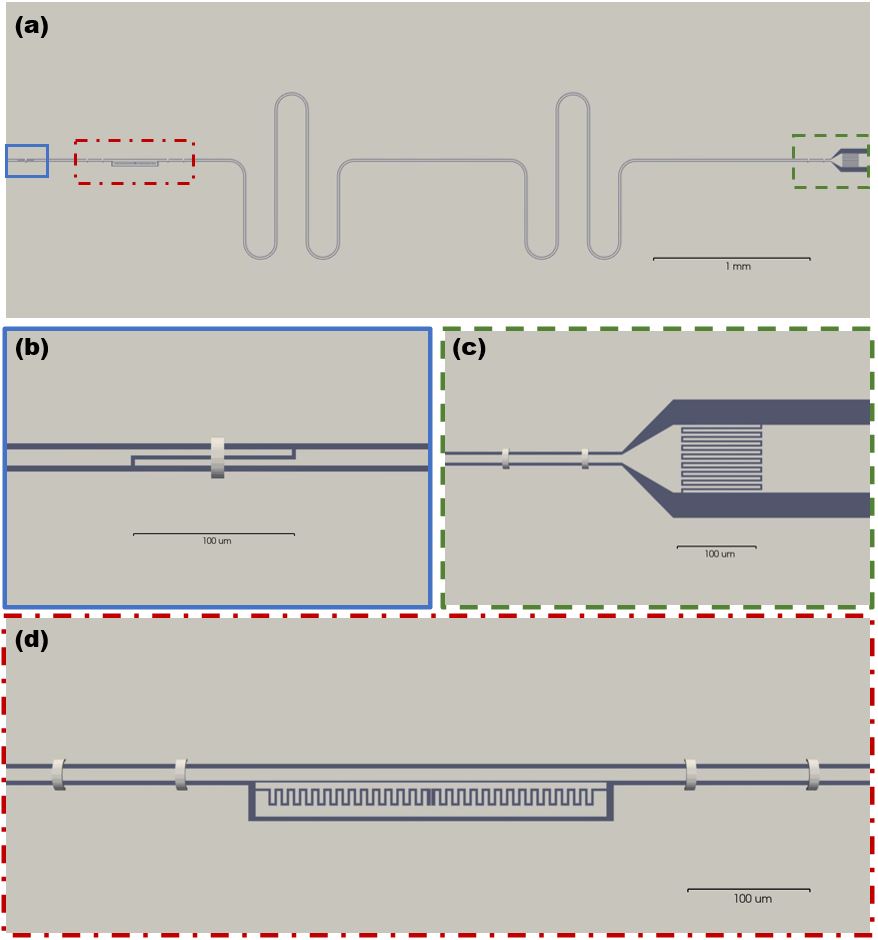}
	\caption{(a) Cavity QED style single photon source geometry. (b) Zoomed in view of the input coupling capacitance, (c) the output coupling capacitance, and (d) the transmon qubit. Airbridges are placed around locations that are asymmetric to minimize the generation of parasitic modes.}
	\label{fig:full_assembled_houck_sps}
\end{figure}

\begin{figure}[t!]
	\centering
	\includegraphics[width=\linewidth]{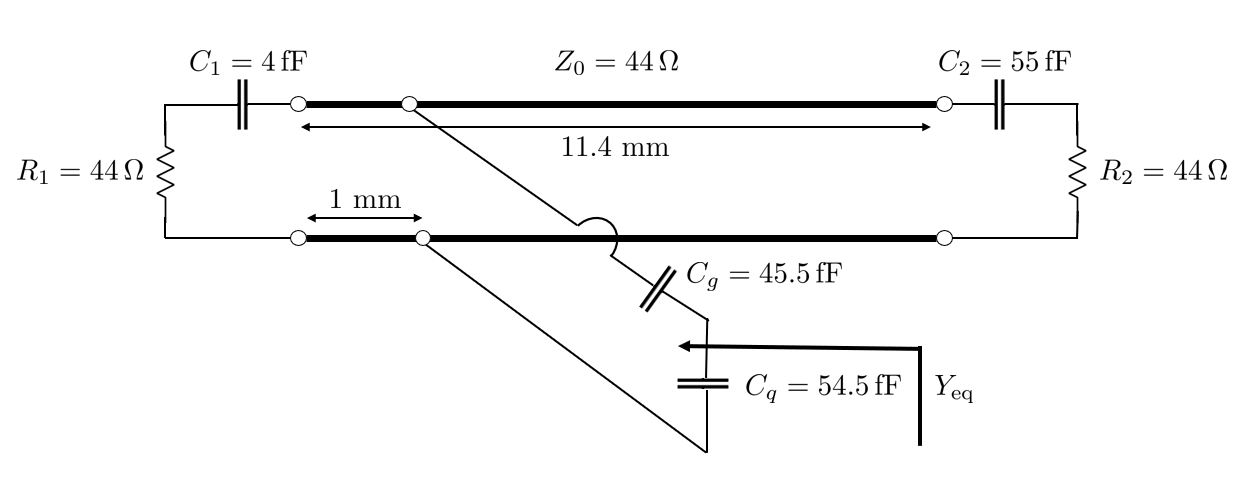}
	\caption{Simplified transmission line model for the cavity QED single photon source shown in Fig. \ref{fig:full_assembled_houck_sps}. The input and output coupling capacitances are represented by $C_1$ and $C_2$, respectively. The resistive terminations $R_1$ and $R_2$ represent the input and output ports the device would be connected to. Other parameters include the coupling capacitance between the transmon and the CPW resonator $C_g$ and the qubit capacitance $C_q$. Non-bold lines in the schematic represent connections with zero electrical length.}
	\label{fig:houck_ckt_model}
\end{figure}

The next device that we will study is a cavity QED style single photon source that was originally presented in \cite{houck2007generating}. Our version of this device is shown in Fig. \ref{fig:full_assembled_houck_sps}. This device consists of a transmon qubit capacitively coupled to a half-wavelength coplanar waveguide (CPW) resonator with resonance frequency of \textapprox5 GHz. The resonator is asymmetrically coupled to two transmission lines through interdigital capacitors shown in Figs. \ref{fig:full_assembled_houck_sps}(b) and \ref{fig:full_assembled_houck_sps}(c). A simple transmission line model of this device from the perspective of the transmon qubit is shown in Fig. \ref{fig:houck_ckt_model}. The circuit parameters are selected to maximize the agreement between the transmission line model and the full-wave results, which are discussed in more detail shortly.

The single photon source is operated by applying a classical microwave drive pulse to the ``input'' line that is coupled to the resonator through the smaller interdigital capacitor to the left of the device. After raising the transmon to its first excited state, the transmon is then allowed to freely evolve in time until it spontaneously emits the excitation predominantly into the CPW resonator due to the tight coupling with the transmon. The photon then leaks out of the resonator predominantly into the ``output'' line that is coupled to the resonator through the larger interdigital capacitor. 

\begin{figure}[t!]
	\centering
	\includegraphics[width=\linewidth]{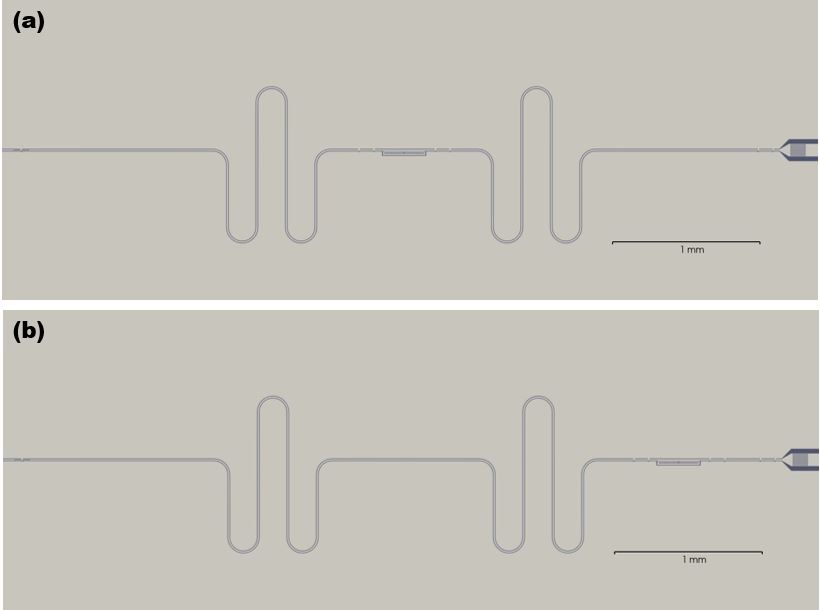}
	\caption{Additional configurations of the cavity QED style single photon source shown in Fig. \ref{fig:full_assembled_houck_sps}(a): (a) middle qubit configuration and (b) output qubit configuration. Every other feature besides the transmon location is identical to the input qubit configuration shown in Fig. \ref{fig:full_assembled_houck_sps}.}
	\label{fig:full_assembled_houck_sps2}
\end{figure}

In our model, the CPW has a signal trace width of 10 $\mu$m and a gap width of 3.75 $\mu$m throughout all of the device except near the output line. At the output line, a signal trace width of 85.5 $\mu$m and a gap width of 32 $\mu$m are used. We assume a silicon substrate and compute the effective permittivity of the lines to be $\epsilon_\mathrm{eff} \approx 6.325$ and the characteristic impedance of the lines to be $44\, \Omega$. As with the model discussed in Section \ref{subsec:waveguide-qed-sps}, we simplify our full-wave model by replacing the thick substrate with a homogeneous background medium with relative permittivity equal to $\epsilon_\mathrm{eff}$ and treat all conductors as being infinitely thin.

The total length of the transmon qubit is 0.288 mm, which is relatively small compared to the wavelength (\textapprox23.86 mm at 5 GHz). As a result, a simple ``point coupling'' approach is often used in developing transmission line and lumped element theoretical models. This approximation can lead to unrealistically fast variations in the computed $T_1$, particularly when the transmon is located near a voltage null in the resonator. To illustrate this, we vary the position of the transmon qubit in our model through three positions. The first is shown in Fig. \ref{fig:full_assembled_houck_sps}(a), which is termed the ``input qubit'' configuration due to the proximity of the qubit to the input transmission line. The two other transmon locations are shown in Fig. \ref{fig:full_assembled_houck_sps2}, which consist of placing the transmon at the middle of the resonator (termed ``middle qubit'' configuration) and near the output line (termed ``output qubit'' configuration). The middle qubit configuration should place the center of the transmon qubit near a voltage null at the first resonant frequency of the CPW resonator.

\begin{figure}[t!]
	\centering
	\includegraphics[width=\linewidth]{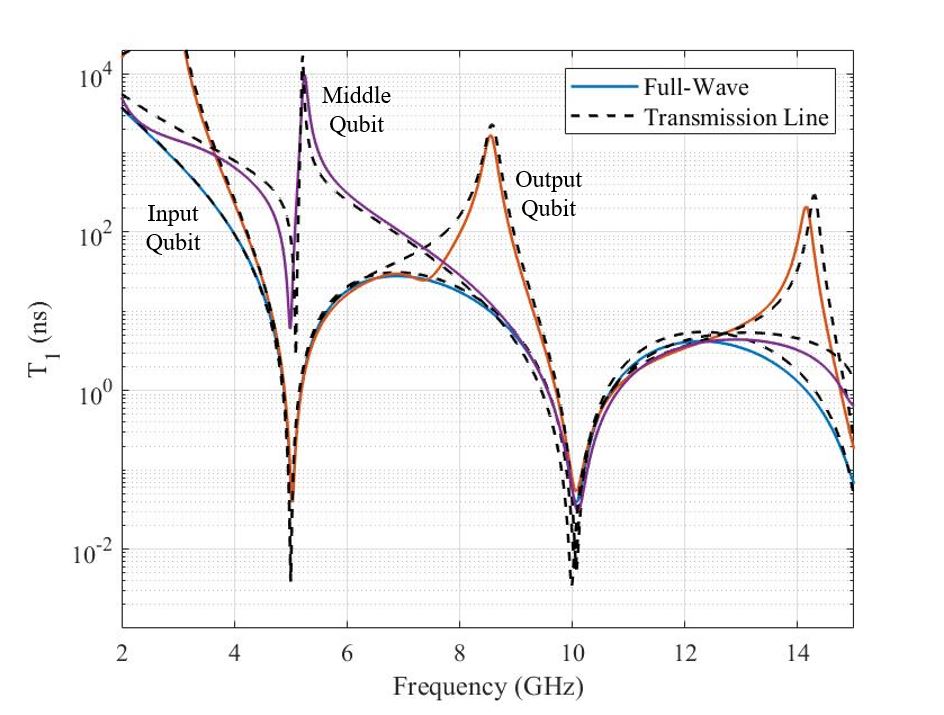}
	\caption{Predictions of the cavity QED style single photon source $T_1$ time for the three qubit locations. Full-wave model results are shown with solid lines and transmission line model results are shown with dashed lines.}
	\label{fig:houck_sps_results}
\end{figure}

The results from our full-wave model for all three transmon configurations are shown in Fig. \ref{fig:houck_sps_results}. Due to the complexity of the device, a lumped element circuit model would lead to significantly incorrect results \cite{houck2008controlling}. Hence, we instead use a transmission line model like that proposed in \cite{houck2008controlling} to compare to our full-wave results in Fig. \ref{fig:houck_sps_results}. We note that the transmission line model parameters are only tuned once, making the only change in the model for the different qubit configurations the location of the qubit. We see that both models do a good job at predicting how the asymmetry of the interdigital capacitors leads to significantly different trends in the $T_1$ time for the input and output qubit configurations. These trends also qualitatively agree very well with the experimental results presented in \cite{houck2008controlling}. 

We also see in Fig. \ref{fig:houck_sps_results} that the middle qubit configuration leads to a complex variation of the $T_1$ time near the first resonance frequency of the CPW resonator. Comparing to the transmission line model, we see that the full-wave results provide a more realistic prediction of how quickly the results can vary as a function of frequency due to the finite spatial extent of the transmon qubit. Accurately understanding the rapid variation of these parameters as a function of frequency has important implications on device design, and can also play an important role in the control dynamics of the qubit \cite{lodahl2015interfacing}. As a result, not ``over-predicting'' the speed of frequency variations can be necessary to make realistic qubit performance predictions.

\subsection{Purcell-Filtered Qubit}
\label{subsec:purcell-filter}
\begin{figure}[t!]
	\centering
	\includegraphics[width=\linewidth]{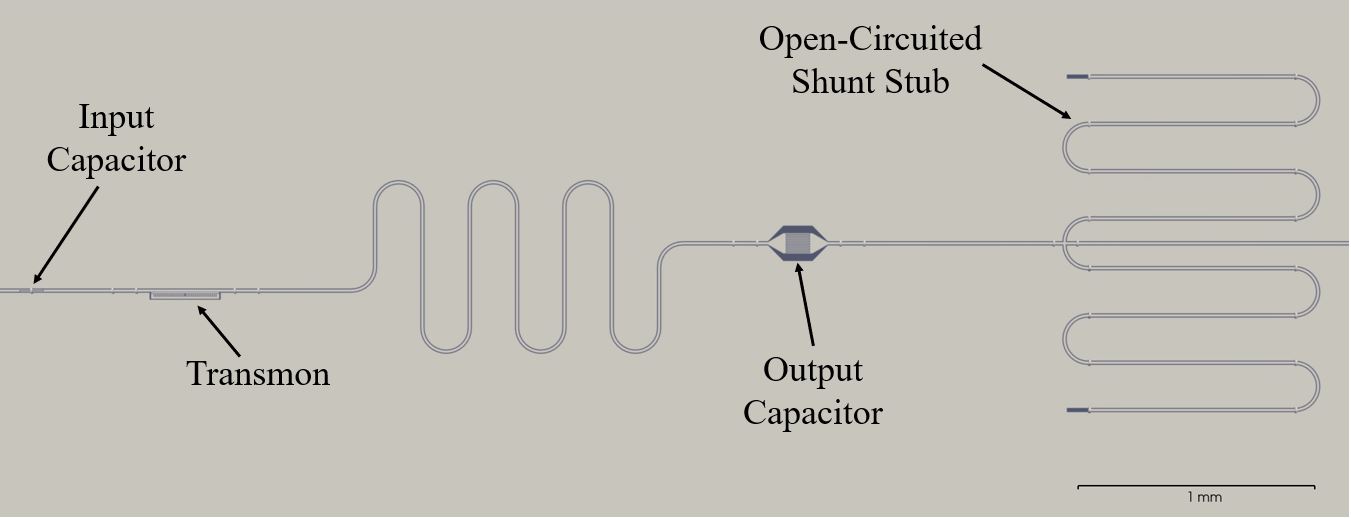}
	\caption{The Purcell-filtered qubit geometry. The input capacitor and transmon are identical to those shown in Figs. \ref{fig:full_assembled_houck_sps}(b) and \ref{fig:full_assembled_houck_sps}(d), respectively. The output capacitor has the same dimensions to that shown in Fig. \ref{fig:full_assembled_houck_sps}(c), but tapers back to the smaller CPW dimensions used throughout the geometry.}
	\label{fig:full_assembled_purcell}
\end{figure}

\begin{figure}[t!]
	\centering
	\includegraphics[width=\linewidth]{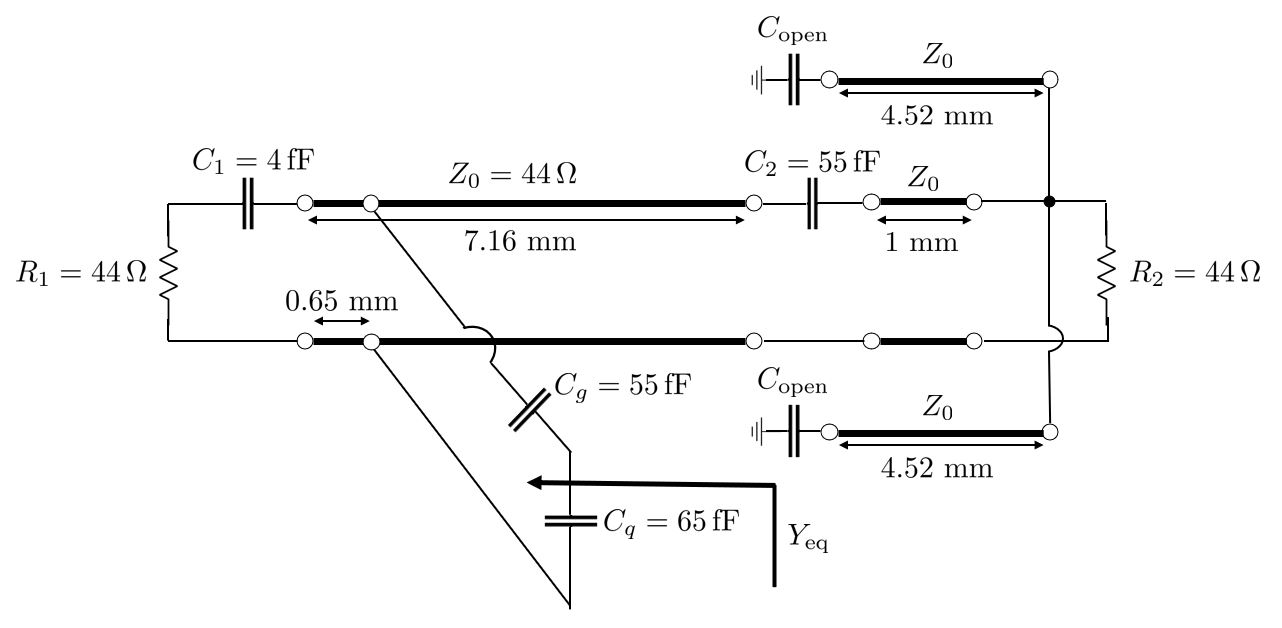}
	\caption{Simplified transmission line model for the Purcell-filtered qubit shown in Fig. \ref{fig:full_assembled_purcell}. Definitions of circuit parameters mirror those in Fig. \ref{fig:houck_ckt_model}. For the model results shown in Fig. \ref{fig:purcell_filter_results}, $C_\mathrm{open} = 0.01$ fF; however, the model results are insensitive to this parameter so long as it is not set to an unrealistically large value (e.g., 100 fF).}
	\label{fig:purcell_ckt_model}
\end{figure}

The final device that we analyze is a qubit with a Purcell filter, similar to the device presented in \cite{reed2010fast}. Our version of this device is shown in Fig. \ref{fig:full_assembled_purcell}, with a simplified transmission line model shown in Fig. \ref{fig:purcell_ckt_model}. This device consists of a transmon capacitively coupled to a CPW resonator in a manner similar to the device discussed in Section \ref{subsec:cavity-qed-sps}. However, the output CPW line that the CPW resonator is coupled to is loaded with two quarter-wavelength open-circuited shunt stubs (where the wavelength here corresponds to the operating frequency of the qubit, not the CPW resonator). These stubs (also known as Purcell filters in this context) have the effect of reducing $\mathrm{Re}\big\{ Z_\mathrm{in} \}$ by ``shorting out'' the output transmission lines at the operating frequency of the qubit, thereby lowering the SER. This helps break the link between the $T_1$ time of the qubit and the CPW resonator's quality factor, which allows the quality factor to be optimized somewhat independently for other considerations related to qubit state readout \cite{reed2010fast}. 

Due to the lack of design information in \cite{reed2010fast}, we largely reuse the features from the device discussed in Section \ref{subsec:cavity-qed-sps}. This includes using the same substrate and effective permittivity, as well as the dimensions of the CPW line, the transmon qubit, and the interdigital coupling capacitors. However, we do adjust the resonant frequencies of the various components to be more in line with those presented in \cite{reed2010fast}. This leads to the CPW resonator the transmon is capacitively coupled to having a resonant frequency of 7.69 GHz and the quarter-wave shunt stubs having resonant frequencies of 6.58 GHz.

\begin{figure}[t!]
	\centering
	\includegraphics[width=\linewidth]{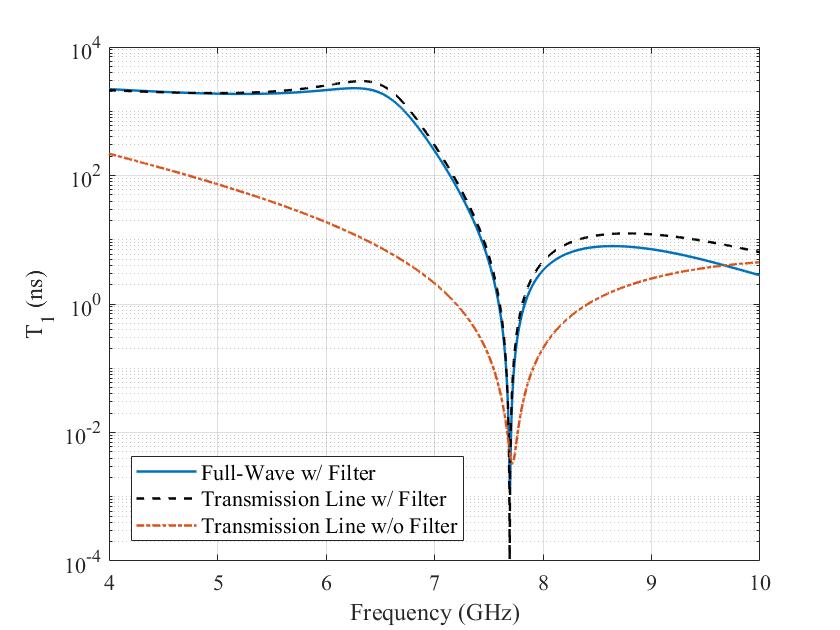}
	\caption{Predictions of the Purell-filtered qubit $T_1$ time. For comparison, results from a transmission line model without the Purcell filter are also shown to demonstrate the efficacy of this simple filter at increasing the qubit $T_1$ time.}
	\label{fig:purcell_filter_results}
\end{figure}

The results from our full-wave and transmission line models for this device are shown in Fig. \ref{fig:purcell_filter_results}, which demonstrates the good agreement achievable between the two models once the parameters of the transmission line model have been optimized. For comparison, we also show the results of a transmission line model without the quarter-wave shunt stubs. This highlights the large increase in $T_1$ time possible even with a relatively simple filtering circuit. These results also agree with the general trends seen in \cite{reed2010fast}, although more differences exist with experimental results due to other sources of decoherence that limited the measured $T_1$ time below that expected from a purely electromagnetic analysis. 

\section{Conclusion}
\label{sec:conclusion}
In this work, we have provided a simple methodology to use the power of full-wave computational electromagnetics tools to predict the SER of transmon qubits in complex architectures. We validated our approach by comparing to various approximate theoretical models that have shown good agreement with experimental results for relatively simple devices. We demonstrated that our full-wave models make more realistic predictions for the variation of the SER as a function of frequency when the transmon is located near rapidly changing spatial fields (e.g., near voltage nulls) by accurately incorporating the spatial extent of the transmon qubit in the model. 

Our full-wave approach can also easily capture non-ideal effects, such as the production of parasitic modes, which can be difficult to accurately incorporate into transmission line or lumped element models. Similarly, our full-wave approach can be easily applied to more complicated devices where developing a simplified lumped element or transmission line model may be impractical. We expect these situations to become increasingly prevalent as the packing density of qubits and supporting control systems continues to rapidly increase in the efforts to make more powerful quantum information processing devices. 

Future work can focus on extending these full-wave modeling approaches to other kinds of circuit QED qubits \cite{gu2017microwave}. There is also a need to more accurately account for other kinds of decoherence in the overall operation of circuit QED systems. Developing field-based approaches that can be reduced to computations involving the classical dyadic Greens' function of the electromagnetic system is a promising approach to incorporate these effects in the analysis of practical systems.

%\appendices
%\input{appendix}

%\section*{Acknowledgment}

%The authors would like to acknowledge useful discussions with Tian Xia, Shu Chen, Thomas Roth, Carlos Salazar-Lazaro and Professor %Scott P. Carney.
%We would like to acknowledge the following funding sources: UIUC CAS, AF Sub RRI PO0539, NSF ECCS 1609195, Ansys Inc PO37497 and the %George and Ann Fisher Professorship.

\ifCLASSOPTIONcaptionsoff
  \newpage
\fi

\bibliographystyle{IEEEtran}
% Put references in BibTeX format in thesisrefs.bib.
\bibliography{./paper_main_bib}


%\newpage



%\vfill


\end{document}